\documentstyle[aps,twocolumn,prl,epsf]{revtex}

\begin{document}

\newcommand{\be}{\begin{equation}}
\newcommand{\ee}{\end{equation}}
\newcommand{\bn}{\begin{eqnarray}}
\newcommand{\en}{\end{eqnarray}}

\draft

\twocolumn[\hsize\textwidth\columnwidth\hsize\csname @twocolumnfalse\endcsname

\title{Metallizing the Mott insulator $TiOCl$ by electron doping: a view
from LDA+DMFT}

\author{L. Craco$^1$, M. S. Laad$^2$ and E. M\"uller-Hartmann$^1$}

\address{$^1$Institut f\"ur Theoretische Physik, Universit\"at zu K\"oln, 
Z\"ulpicher Strasse, 50937 K\"oln, Germany  \\
$^2$Department of Physics, Loughborough University, LE11 3TU, UK}

\date{\today}
\maketitle

\widetext

\begin{abstract}
Based on recent experiments, we describe the Mott insulating, but 
undimerized state of $TiOCl$ using the local-density approximation 
combined with multi-orbital dynamical mean field theory (LDA+DMFT) 
for this $3d^{1}$ system.Good agreement with the high estimated value 
of the superexchange is obtained. The possibility of an electron-doped 
insulator-metal transition in $TiOCl$ is investigated in this scheme 
and a Mott-Hubbard transition with a jump in carrier density {\it inside} 
the metallic state is found.  Clear, co-incident, discontinuous changes 
in orbital occupations are observed, showing that such a transition 
would involve strong, multi-orbital correlations. These results call 
for studies on suitably intercalated $TiOCl$ that may induce 
metallization and, possibly, unconventional superconductivity.
\end{abstract}

\pacs{PACS numbers: 
71.27.+a, %Strongly correlated electron systems; heavy fermions 
71.15.Mb, %Density functional theory, local density approximation, ...
79.60.-i  %Photoemission and photoelectron spectra 
}

]

\narrowtext

Titanium-based oxide systems have recently been found to exhibit 
varied physical properties. Being electron-doped analogues ($3d^{1}$) 
of the well-known high-$T_{c}$ superconductors, they have been studied with 
a view to search for superconductivity. Titanium-oxychloride ($TiOCl$) is 
of special interest in this context~\cite{[1]}: a layered material undergoing 
spin-Peierls dimerization at low $T$. Unlike other candidate spin-Peierls 
systems like $CuGeO_{3}$, the superexchange scale $J\simeq 660~K$ 
in $TiOCl$ is almost an order-of-magnitude larger. Additionally, the fact 
that the single $d$ electron resides in the $t_{2g}$ orbitals might suggest 
the absence of Jahn-Teller effects, implying weakened tendency for doped 
carriers to localize. It has nevertheless proved to be difficult to dope 
$TiOCl$ using traditional techniques. However, structurally similar materials 
such as $Li$-intercalated $ZrNCl$ are known to undergo insulator-metal 
transition and superconductivity~\cite{[2]}, leaving the issue open for 
$TiOCl$.

Early observation of spin-gap generation in spin susceptibility showing a 
kink at $94~K$ and an exponential drop below $T_{d}=66~K$, accompanied by 
a doubling in the unit cell along $b$ suggested a spin-Peierls dimerized 
state in $TiOCl$. That this may in fact involve coupled spin, orbital and 
lattice degrees of freedom is suggested by large phonon anomalies~\cite{[3]} 
and by $T$ dependent $g$-factors and ESR linewidths~\cite{[4]}. This should 
not be too surprising: in fact, the magnetism of early TMOs is known to be 
sensitively determined by the above factors~\cite{[5]}. Indeed, Seidel 
{\it et al.}~\cite{[3]} and Saha-Dasgupta {\it et al.}~\cite{[6]} have 
recently studied the electronic structure of $TiOCl$ using the LDA+U method: 
in contrast to what is expected from a cubic crystal field, the low-symmetry 
crystal field in $TiOCl$ splits the $t_{2g}$ orbitals into a lower lying 
$d_{xy}$ and higher-lying $d_{yz,zx}$ orbitals. The lone electron in $TiOCl$ 
would then occupy the lowest $d_{xy}$ orbital at low-$T$. However, the 
higher-lying $d_{yz,zx}$ orbitals are only about $0.2~eV$ above the lowest 
one, and, in a situation where the LDA bandwidth is about $2.5~eV$, strong 
inter-orbital mixing can be expected. How the system manages to achieve a 
spin-Peierls insulating state when moderately strong, multi-orbital Coulomb 
interactions in the $d$ shell are switched on is a major question for
{\it ab initio} theoretical approaches.  

In this letter, we aim to address precisely this issue in part. Starting with 
the LDA calculation for the actual structure of $TiOCl$, we explicitly treat
the dynamical effects of strong, multi-orbital electronic correlations in the
$3d^{1}$ case using multi-orbital dynamical mean field theory (DMFT). To solve
the DMFT equations, we employ the multi-orbital iterated perturbation theory
(IPT) used successfully earlier~\cite{[7]}. This is in fact necessary: the 
LDA generically gives metallic states owing to neglect of strong electron 
correlations, while LDA+U, which does give the correct {\it ground} state(s) 
for insulating systems, generically overestimates localization, leading to 
charge gap(s) too large compared to experiment. Moreover, being ground state 
theories, both are incapable of accessing the ubiquitous dynamical spectral 
weight transfers in response to various small perturbations; the importance 
of such aspects is known to be crucial, e.g, in insulator-metal transitions.  
This characteristic underlying the ill-understood response of correlated 
systems can however be understood within DMFT and its extensions.  

\begin{figure}[htb]
\epsfxsize=3.in
\epsffile{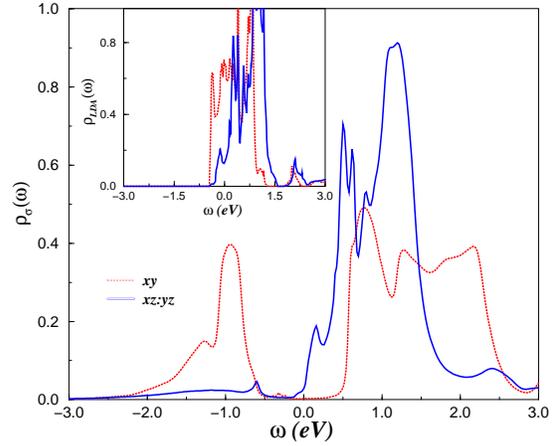}
\caption{%%%(Color online) 
LDA+DMFT and LDA (inset) partial densities 
of states (DOS) for $TiOCl$. Dotted (red) line denotes the DOS for 
the non-degenerate ground-state $(xy)$ orbital, the full (blue) line 
denotes the DOS for the higher (two-fold degenerate) lying $t_{2g}$ 
orbitals. Higher-lying $e_{g}$ and $O$-2p bands are not shown.}
\label{fig1}
\end{figure}

$TiOCl$ crystallizes in the $Pmmn$ space group at high-$T$, consisting of 
bilayers of $Ti^{3+}$ and $O^{2-}$ parallel to the {\it ab} plane separated 
by $Cl^{-}$ layers. In a perfect octahedral environment the crystal field 
would split the atomic $d$ levels into three-fold degenerate ($t_{2g}$) and 
doubly degenerate $e_{g}$ levels. However, the basic $TiCl_{2}O_{4}$ unit 
in $TiOCl$ is distorted, and the low-symmetry crystal field in the real 
structure splits the $t_{2g}$ degeneracy, with the lowest $d_{xy}$ orbital 
separated from slightly higher lying $d_{yz,zx}$ orbitals. The inset of 
Fig.~\ref{fig1} shows the one-electron density of states (DOS) obtained 
using density-functional calculations in the LDA. As expected, LDA does 
not give an insulating solution. We now show how LDA+DMFT reconciles this 
problem, yielding not only the correct ground state, but also a 
quantitative picture for one-electron (hole) excitation spectrum. The 
LDA Hamiltonian is 
%\be
$H_{0}=\sum_{k,\alpha\beta}\epsilon_{k}^{\alpha\beta}
c_{k\alpha\sigma}^{\dag}c_{k\beta\sigma} 
+ \sum_{i,\alpha\sigma}\epsilon_{i\alpha\sigma}^{0}n_{i\alpha\sigma},$
%\ee
where 
$\epsilon_{i\alpha\sigma}^{0}=\epsilon_{i\alpha\sigma}-U(n_{\alpha\bar\sigma}
-1/2)+\frac{J_{H}}{2}\sigma(n_{\alpha\sigma}-1)$ with $U$ and $J_{H}$ as 
defined below. The second term above takes care of avoiding the 
double-counting of interactions already treated on the average by LDA.

%%%%%The above discussion shows the multi-orbital character of the system. 
We account for strong multi-orbital correlations in $TiOCl$ by  
choosing~\cite{[6]} the intra-orbital $U$, the local Hund coupling 
$J_{H}$ and the inter-orbital $U'\simeq U-2J_H$. We stress the importance 
of including $U'$: indeed, neglecting $U'$ would imply that an electron 
hopping from site $i$ to its neighbor 
(avoiding the same lowest orbital to escape the penalty of $U$) 
would occupy the slightly higher-lying $d_{yz,zx}$ orbitals.  
With $U'=0$, and a one-electron band-width of $2.0~eV$, this electron 
(and the hole it leaves behind on the site $i$) would hop freely, always 
giving a {\it metallic} state.  The full many-body Hamiltonian is
%\be
$H=H_{0}+U\sum_{i,\alpha}n_{i\alpha\uparrow}n_{i\alpha\downarrow} 
+ \sum_{i\alpha\beta\sigma\sigma'}U_{\alpha\beta}^{\sigma\sigma'}
n_{i\alpha\sigma}n_{i\beta\sigma'}$.
%\ee
For early TMOs, it is sufficient to consider only the $t_{2g}$ bands in 
the LDA+DMFT scheme~\cite{[10]}. Due to $t_{2g}-e_{g}$ and $t_{2g}-2p$ 
hybridization in the real crystal structure of $TiOCl$, we include the 
TM-$e_{g}$ and $O$-2p parts (of the DOS) with ``$t_{2g}$'' symmetry in 
the unperturbed band DOS. Further, in the paramagnetic/ferro-orbital
situation (see below) applicable to $TiOCl$ in the undimerized state, 
we have, $G_{\alpha\beta\sigma\sigma'}=
\delta_{\alpha\beta}\delta_{\sigma\sigma'}G_{\alpha\sigma}(\omega)$ and
$\Sigma_{\alpha\beta\sigma\sigma'}(\omega)=
\delta_{\alpha\beta}\delta_{\sigma\sigma'}\Sigma_{\alpha\sigma}(\omega)$. 

The DMFT solution in the $t_{2g}$ sector involves $(i)$ replacing the 
lattice model by a multi-orbital, asymmetric Anderson impurity model, 
along with $(ii)$ a selfconsistency condition which requires the impurity 
propagator to coincide with the local Green's function of the lattice, 
given by
%\be
$G_{\alpha}(\omega)=\frac{1}{V_{B}}\int d^{3}k[\frac{1}{(\omega+\mu)1
-H_{LDA}^{0}-\Sigma_{\alpha}(\omega)}]_\alpha$.
%\ee
Using the locality of $\Sigma_{\alpha}$ in $d=\infty$, we have 
$G_{\alpha}(\omega)=G_{\alpha}^{0}(\omega-\Sigma_{\alpha}(\omega))$. Further, 
$U',J_{H}$ scatter electrons between the $d_{xy}$ and $d_{yz,zx}$ bands, so 
only the total number, $\sum_{\alpha}n_{t_{2g}^\alpha}$, is conserved in 
accord with Luttinger's theorem. In the calculations, strict compliance with 
the Friedel-Luttinger sum rule (guaranteeing correct low-energy behavior) 
and with the first three moments of the spectral function (guaranteeing 
correct high-energy behavior) is maintained~\cite{[11]}.

To solve the multi-orbital single-impurity problem, we use the multi-orbital 
(MO) IPT, suitably generalized for arbitrary band-filling and temperatures.  
This method has been successfully used to solve a host of related 
problems~\cite{[11]}, and we refer the reader to the established literature 
for the details. We input the LDA DOS ($\rho_{\alpha}^{0}(\epsilon)$) into 
the DMFT(MO-IPT). In a multi-orbital system, DMFT leads to two effects: 
first, the LDA parameters like crystal field splitting are renormalized in 
non-trivial ways from their bare values, and may even change 
sign~\cite{[7]}. These effects result from the multi-orbital Hartree 
shifts, which correspond to effects captured by LDA+U. Secondly, and more 
importantly, DMFT introduces non-trivial effects stemming from the 
{\it dynamical} nature of the strong electronic correlations. Namely, these 
processes lead to large transfer of spectral weight across large energy 
scales in response to small changes in the {\it bare} electronic structure, 
a characteristic lying at the heart of the anomalous responses of correlated 
systems~\cite{[5]}.

Let us now present our results. In Fig.~\ref{fig1}, we show the 
orbitally-resolved spectral functions for the $d_{xy}$ (dotted line) and 
the $d_{yz,zx}$ (solid) parts for $U=3.3~eV$, with $J_{H}=1.0~eV$ and 
$U'=1.3~eV$. As discussed above, DMFT results in non-trivial 
renormalization of the LDA crystal field splitting, 
$\Delta_{t_{2g}}=\epsilon_{xy}-\epsilon_{\alpha}$ $(\alpha \equiv yz,zx)$ 
from $-0.45~eV$ to 
$0.12~eV$ in LDA+DMFT. More important, a clear charge gap showing the 
Mott-Hubbard character of the insulator is selfconsistently generated (the 
very small finite DOS at $\omega=0$ is a result of finite $T \approx 100~K$ 
effects). The drastic change in the lineshape attests to the explicit 
consideration of dynamical multi-orbital correlations. Photoemission 
spectroscopy should be able to confirm this interpretation: if true, a 
lower Hubbard band feature around $-1.0~eV$ with an asymmetric profile 
should be observable. Also XAS should show a shoulder feature at 
low-energies {\it above} the Mott gap, and the XAS weight should be much 
higher than in PES.   

An important result directly gleaned from Fig.~\ref{fig1} is that 
$n_{d_{xy}}:n_{d_{\alpha}}=7:3$, clearly showing that the $d_{xy}$ orbital 
is dominantly populated in the ground state (however, with non-negligible 
inter-orbital fluctuations). This is exactly the condition needed to 
generate one-dimensional spin chains along $b$, as observed, and is 
consistent with the LDA+U result. Using the renormalized value of 
$\Delta_{t_{2g}}=0.12~eV$ along with the hopping matrix elements 
$t_{\alpha\beta}$~\cite{[6]} and $U,U',J_{H}$ as given above, the 
effective superexchange constant between nearest neighbor ($S=1/2$) spins 
on $Ti^{4+}$ ions along $b$ can be estimated. Following~\cite{[12]}, 
we get $J\simeq 0.06~meV \cong 720~K$, in nice agreement with the $dc$ 
susceptibility data. This is the dominant contribution to the superexchange, 
and we found that further-neighbor couplings are negligible on the scale 
of $J$. Given this observation, the spin-Peierls instability observed at 
$T_{SP}=66~K$ is almost certainly caused by coupling to lattice degrees 
of freedom, and {\it not} by competing, appreciable next-n.n 
interactions~\cite{[6]}.  
The derivation of this instability is likely to involve low-dimensional 
effects close to and above $T_{SP}$ and is out of scope of {\it any}
mean-field theory. However, we stress that the SP transition arises as 
an (lattice coupling induced spontaneous breaking of translational symmetry 
along $b$) instability of the (high-$T$) Mott insulator derived above. 
This can be verified, for example, by studying the evolution of the $dc$ 
spin susceptibility upon adding a small number of non-magnetic impurities 
at $Ti$ sites: in analogy to $V_{1-x}Cr_{x}O_{2}$~\cite{[5]}, we predict a 
Curie component growing with the impurity concentration, which should not 
be observed if the I-phase is a band-Peierls insulator. 

We now consider the effect of pressure and carrier doping: at the outset, 
we clarify that attempts to dope $TiOCl$ have not succeeded to date. On the 
other hand, the fact that related systems like $Li_xZrNCl$ and
$M_xHfNCl~(M=Li, Na)$~\cite{[2],[13]} have been shown to become metallic 
and superconducting upon {\it electron} doping suggests that attempts to use 
similar strategies in $TiOCl$ may work~\cite{also}. With this caveat, 
we have studied the issue of carrier density driven I-M (Mott) transition(s) 
in our LDA+DMFT study.  

{\it Hole doping (Case 1)}: We have repeated the LDA+DMFT calculation for 
hole-doping ($x$), corresponding to band-filling $n_{t_{2g}}=1-x$. No 
instability to the metallic state is found up to $x=0.1$. We also performed 
the calculations by changing $\Delta_{t_{2g}}$ in small trial steps (beginning 
with its value $\Delta_{t_{2g}}^{I}$) with constant $n_{t_{2g}}=1.0$, as done
earlier in another context~\cite{[7]} to mimic pressure driven changes, and 
found no I-M transition. Notice that hole doping will predominantly involve 
changes in the lowest, $d_{xy}$ band. The absence of hole-doping driven I-M 
transition is unusual and contrary to naive expectations from ($d=\infty$) 
studies of the doped Hubbard model, and its absence under pressure is 
consistent with high-pressure studies which have hitherto found no I-M 
transition up to $P=60~kbar$~\cite{[1]}.  

{\it Electron doping (Case 2)}: A dramatically different outcome is 
realized with electron ($n_{t_{2g}}=1+x$) doping. A nearly first-order 
I-M transition with rapid change in the carrier density around 
$n_{t_{2g}}=1.9$ is clearly seen in the LDA+DMFT results. In $TiOCl$, 
$e^{-}$-doping necessarily involves doping the $d_\alpha$ bands, in 
contrast to hole doping, which affects mainly the $d_{xy}$ band.
 
This finding has a natural interpretation in a multi-orbital Mott-Hubbard 
picture. Beginning with {\it Case 2},
\begin{figure}[htb]
\epsfxsize=3.in
\epsffile{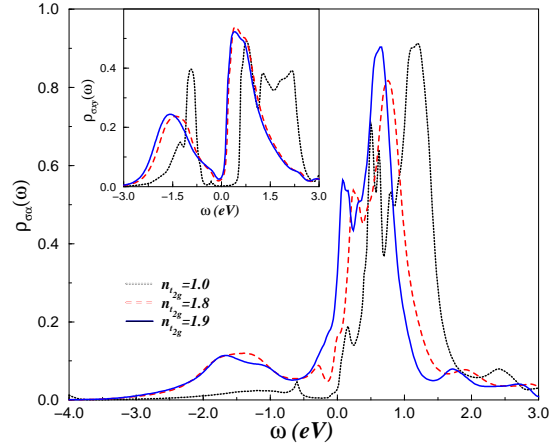}
\caption{%%%%(Color online) 
$t_{2g}$ partial DOS for the $3d_{xz:yz}$ and $3d_{xy}$ (inset) orbital 
for different values of the total electron number.}
\label{fig2}
\end{figure}
\hspace{-0.4cm}{\it electrons} added to the Mott 
insulator (in our calculation, we change the electron number to 
$n_{t_{2g}}=1+x$ and search for the I-M instability starting from the Mott
insulator found above) prefer to occupy the higher ($d_\alpha$) orbitals 
to escape the large ($U$) cost for doubly occupied $d_{xy}$ orbitals 
($U'<U$). In this situation, the renormalized $\Delta_{t_{2g}}=0.12$ (MI) 
changes upon doping by an amount related to $U'$ and to the self-consistently 
calculated average occupations $n_{xy,\alpha}$ of each $t_{2g}$ orbitals.  
This in turn changes $\Delta_{t_{2g}}$. More importantly, the 
{\it dynamical} multi-orbital correlations ($U,U',J_{H}$) react drastically 
to this small change in $\Delta_{t_{2g}}$, transferring spectral weight 
over large energy scales from high- to low energies, and driving a Mott
I-M transition at a critical $x$, as shown in Fig.~\ref{fig2}. We clearly 
observe that {\it only} the $d_{\alpha}$ bands show metallic behavior; the 
$d_{xy}$ DOS still represents almost insulating behavior. This is a direct 
consequence of the fact that it is the $d_{xy}$ orbital which is 
predominantly occupied in the MI. In this limit, the multi-orbital 
Hartree shift for the $d_\alpha$ orbitals pulls the $d_\alpha$ 
states lower, populating them more (given the constraint), driving more 
spectral weight transfer from high- to low energy (MO-DMFT), 
selfconsistently generating a metallic state. The renormalized value of
$\Delta_{t_{2g}}^{M}=-0.011$ in the metallic ($n_{t2g}=1.9$) state should 
imply a ``melting'' of the spin dimerization observed in the I phase.  More 
importantly, it also implies a drastic change in occupation of the various 
$t_{2g}$ orbitals, corresponding to the sign-change in $\Delta_{t_{2g}}$.  
The situation seems to be similar to the one encountered in 
$VO_{2}$~\cite{[5]}. In the Ginsburg-Landau picture of the MIT, $
\Delta_{t_{2g}}$ plays the role of an external field in the orbital sector, 
which drive changes in orbital occupations, leading to the MIT 
via spectral weight transfer (corresponding to the second solution of 
the DMFT equations) beyond a critical $x$-dependent value. 

Clearly, the case of hole doping is very different: holes will predominantly 
occupy the lowest $d_{xy}$ orbitals. The multi-orbital Hartree shift is 
now ineffective.  Since the $d_{xy}$ DOS shows a large ($\simeq 1.0~eV$) 
gap in the $I$-phase, small changes in $\epsilon_{xy}$ with $x$, along with
the absence of multi-orbital shifts (which produced the lowering of $\alpha$ 
orbitals in {\it Case 2}) are not sufficient to drive an I-M transition 
in {\it Case 1}. In reality, given the quasi-1D character of the hopping 
involving $d_{xy}$ orbitals, the doped holes would be immediately localized 
by disorder: this is an effect which will operate at least above a certain 
temperature scale where the system would really behave one-dimensionally. 
At low-$T$ the above mechanism may be operative in reality.

\begin{figure}[htb]
\epsfxsize=3.in
\epsffile{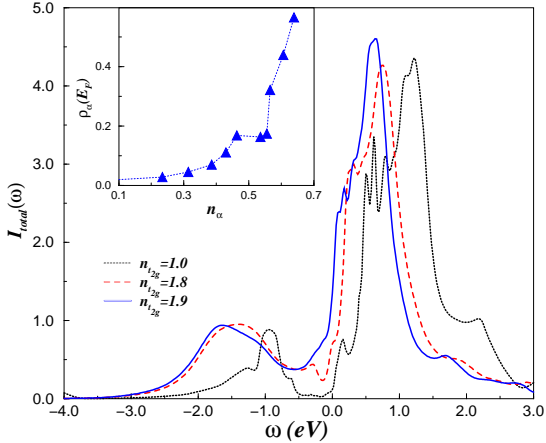}
\caption{%%%(Color online) 
Total one-electron spectra for different values 
of the total electron number. The inset shows $3d_{xz:yz}$ DOS at $E_{F}$ 
as a function of the the corresponding orbital occupation.} 
\label{fig3}
\end{figure}

In Fig.~\ref{fig3}, we show the total spectra in the insulating and metallic 
phases. We clearly see the low energy pseudogap in the DOS for the 
bad-metallic case ($n_{t_{2g}}<n_{cr}=1.8$) and an abrupt change to a much 
more itinerant (large DOS at low energy) metal for $n_{t_{2g}}>n_{cr}$. Large 
spectral weight transfer from high energies to the Fermi level as a function 
of electron doping is also clearly visible. As remarked earlier, such a 
dynamical spectral weight driven I-M transition is out of scope of LDA or 
LDA+U methods, and can only be reliably studied within LDA+DMFT. To further 
explore the non-rigid-band scenario across the insulator-metal transition, 
we show the change in $\rho_{\alpha}(E_{F})$ as a function of $n_{\alpha}$ 
(inset of Fig.~\ref{fig3}) for our chosen parameter set. Clearly, 
$\rho_{\alpha}(E_{F})$ jumps 
sharply from $0.17$ to $0.31$ around $n_{\alpha}=0.56$, indicating an 
unconventional electron-doped first order transition. Our results imply that 
the metallic state of $TiOCl$ is accompanied by an jump in the itinerant 
carrier density at a critical value of the $n_{\alpha}$ orbital occupation 
close to the I-M transition.  Based on our calculation, we tentatively 
 suggest that suitably intercalated (electron doped) $TiOCl$ may
also exhibit unconventional superconductivity upon metallization.
Further, given the multi-band character of the metallic state found above, 
multi-band (spin singlet for $J_{H}<\Delta_{t_{2g}}$ or more probably spin 
triplet for $J_{H}>\Delta_{t_{2g}}$) superconductivity may be realized 
in such systems.  This issue is out of scope of this paper, however. 

To conclude, we have studied the insulating state in $TiOCl$ using the 
ab-initio LDA+DMFT. In good agreement with the $dc$ susceptibility 
measurement, we found that the nearest neighbor superexchange, 
$J\simeq 720~K$ (it is $660~K$ in the experiment). A consistent derivation 
of the spin-Peierls instability from the Mott insulator requires inclusion 
of additional (non-local?) coupling to the lattice, and is planned for the 
future. In accordance with pressure studies, we found no I-M transition 
occuring under pressure in $TiOCl$. Building on this agreement with various 
basic obervations in the I-phase of $TiOCl$, we suggest that electron-doping 
the Mott insulator (maybe achievable by intercalation with suitable species) 
would drive a first-order Mott-Hubbard transition to a correlated metallic 
state. Our results show that if this transition be first order, it will 
be accompanied by discontinuous changes in orbital occupations, a feature 
which is also seen in increasing variety of correlated, multi-orbital 
systems undergoing Mott transitions. Our work calls for studies on 
suitably intercalated $TiOCl$ to look for possible routes to metallization 
and unconventional superconductivity. 
  
%\acknowledgments
The authors are indebted to T. Saha-Dasgupta and R. Valenti for providing
the LDA DOS, and would also like to acknowledge M. Gr\"uninger for 
valuable discussions.
The work of LC was carried out under the auspices of the 
Sonderforschungsbereich 608 of the Deutsche Forschungsgemeinschaft.
MSL acknowledges financial support from the EPSRC (UK).


\begin{references}
\bibitem{[10]} The $O$-2p bands and the TM $e_{g}$ bands are well 
separated from the TM $e_{g}$ bands in early TMO.
\end{references}
\end{document}